\documentclass[10pt,journal,compsoc]{IEEEtran}
\usepackage{algpseudocode}

\usepackage{graphicx}
\usepackage{float}
\usepackage{array}
\usepackage{tabularx}
\usepackage{balance}
\usepackage{marvosym}
\usepackage{amssymb}
\usepackage{amsmath}
\usepackage{multirow}
\usepackage{array}
\usepackage{enumitem}
\usepackage{booktabs} 
\usepackage[ruled,vlined]{algorithm2e}
\usepackage{amsfonts}
\usepackage{graphicx}
\usepackage[caption=false]{subfig}
\usepackage{textcomp}
\usepackage[dvipsnames, svgnames, x11names]{xcolor} 
\usepackage{pifont}
\usepackage{url}
\usepackage{bm}
\usepackage{makecell}
\usepackage{adjustbox}
\usepackage{dsfont}

\usepackage{overpic}
\usepackage{algpseudocode}

\usepackage{enumitem}
\usepackage{balance}

\definecolor{lightblue}{RGB}{0, 134, 179}
\usepackage[colorlinks,linkcolor={black},citecolor={black},urlcolor={black}]{hyperref}

\makeatletter

\newcommand{\Rmnum}[1]{\expandafter\@slowromancap\romannumeral #1@}
\makeatother

\begin{document}

\title{An Empirical Study of Training ID-Agnostic Multi-modal Sequential Recommenders}

\author{Youhua Li, Hanwen Du, Yongxin Ni, Yuanqi He, Junchen Fu, Xiangyan Liu, Qi Guo
\IEEEcompsocitemizethanks{
\IEEEcompsocthanksitem 
Youhua Li is with ShanghaiTech University, who is also with the Institute of Computing Technology, Chinese Academy of Sciences, Beijing, China, and Shanghai Innovation Center for Processor Technologies (SHIC).\protect\\
Email: liyh5@shanghaitech.edu.cn
\IEEEcompsocthanksitem
Qi Guo is with State Key Lab of Processors, Institute of Computing Technology, Chinese Academy of Sciences, Beijing, China.\protect\\
Email: guoqi@ict.ac.cn
}
}

\markboth{IEEE TRANSACTIONS ON KNOWLEDGE AND DATA ENGINEERING}%
{Shell \MakeLowercase{\textit{et al.}}: Bare Demo of IEEEtran.cls for Computer Society Journals}

\IEEEtitleabstractindextext{%
\begin{abstract}Sequential Recommendation (SR) aims to predict future user-item interactions based on historical interactions. While many SR approaches concentrate on user IDs and item IDs, the human perception of the world through multi-modal signals, like text and images, has inspired researchers to delve into constructing SR from multi-modal information without using IDs. However, the complexity of multi-modal learning manifests in diverse feature extractors, fusion methods, and pre-trained models. Consequently, designing a simple and universal \textbf{M}ulti-\textbf{M}odal \textbf{S}equential \textbf{R}ecommendation (\textbf{MMSR}) framework remains a formidable challenge. We systematically summarize the existing multi-modal related SR methods and distill the essence into four core components: visual encoder, text encoder, multimodal fusion module, and sequential architecture. Along these dimensions, we dissect the model designs, and answer the following sub-questions: First, we explore how to construct MMSR from scratch, ensuring its performance either on par with or exceeds existing SR methods without complex techniques. Second, we examine if MMSR can benefit from existing multi-modal pre-training paradigms. Third, we assess MMSR's capability in tackling common challenges like cold start and domain transferring. Our experiment results across four real-world recommendation scenarios demonstrate the great potential ID-agnostic multi-modal sequential recommendation. Our framework can be found at: https://github.com/MMSR23/MMSR.
\end{abstract}
		
}

\maketitle

\IEEEdisplaynontitleabstractindextext

%
\IEEEpeerreviewmaketitle	

\IEEEraisesectionheading{\section{Introduction}\label{sec:introduction}}
 
\IEEEPARstart{S}{equential} Recommendation (SR) leverages users' past interactions to recommend their next item of interest. Its implementation in online environments, like e-commerce platforms and streaming media websites, has significantly expanded the range of options available to customers and has attracted increasing attention in the research community. One of the primary challenges in developing SR models is to generate high-quality user and item representations that can enhance recommendation performances.

\begin{figure}[t]
	\begin{center}
    \includegraphics[width=1.0\linewidth]{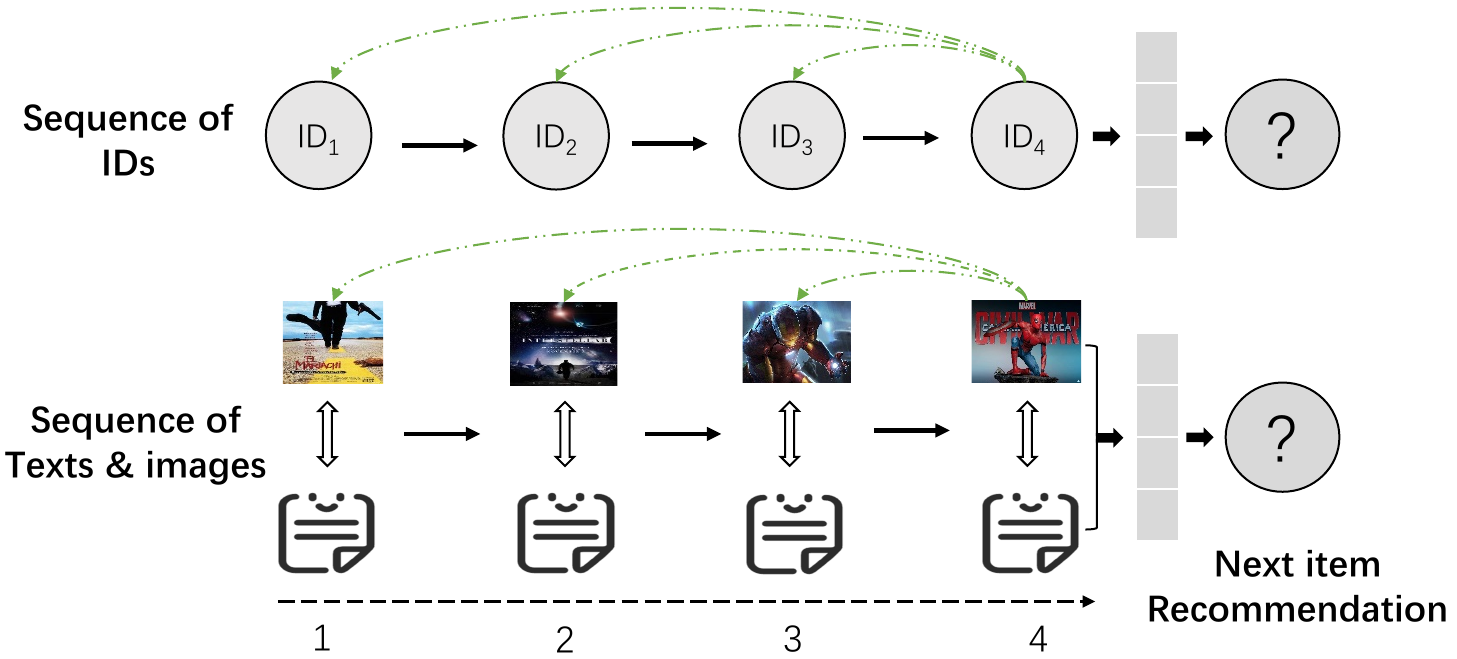}
    \end{center}
    \caption{Sequential recommendation models the historical interactions between users and items and predicts the next item. Up: \textbf{IDSR}. Every item is represented by an ID Embedding. Bottom: \textbf{MMSR}. Every item is represented by its multi-modality information, such as covers and titles. }
    \label{recommendation}
\end{figure}

In mainstream SR scenarios, a user can be represented by all the items that this user has acted upon, with each item encoded as a unique identifier (ID for short), as is shown in the upper part of Figure \ref{recommendation}.
To provide recommendation, the prevailing paradigm in current community is to maintain a learnable embedding matrix for all item IDs, and item representation in sequences can map as a latent vector by looking up its corresponding ID, we summarize this paradigm as ID-based  Sequential Recommendation (\textbf{IDSR}).
For example, GRU4Rec~\cite{hidasi2015session} adopts Recurrent Neural Network (RNN) as the sequence encoder, NextItNet~\cite{yuan2019simple} adopts Convolutional Neural Network (CNN) as the sequence encoder, SASRec~\cite{kang2018self} adopts unidirectional Transformer~\cite{vaswani2017attention} as the sequence encoder, BERT4Rec~\cite{sun2019bert4rec} adopts bidirectional Transformer~\cite{devlin2018bert} as the sequence encoder.
Although straightforward and easy to implement, IDSR has several significant limitations. First, ID representations learned from one domain cannot be transferred across platforms, since the item vocabularies from two non-overlapping domains cannot be shared \cite{kang2020learning, yuan2023go}. Second, IDSR can only learn the representations from the item IDs, leading to recommend cold items from highly skewed data distribution intractable \cite{wei2021contrastive,zhu2021learning}.
Additionally, IDSR's inability to leverage advancements in other fields such as Natural Language Processing (NLP) and Computer Vision (CV) restricts the development of large-scale general model applications within recommender systems, unlike in NLP and CV domains.

To address the cold start problem and enhance the quality of recommendations, numerous Sequential Recommendation (SR) studies \cite{zhang2019feature,xie2022decoupled, MMGCN,zhou2020s3,rashed2022carca} have integrated (multi-)modality information, such as text and images, as auxiliary features. These features aim to construct more comprehensive item representations by combining them with ID-based representations. Despite this integration, these models remain dependent on IDs, limiting their applicability across different domains. This raises a critical question: Are user and item IDs indispensable for recommender systems? We argue that the modality or multi-modality information of an item, such as covers and titles, can serve as unique and distinguishable representations, as illustrated in Figure \ref{recommendation} (below). As an alternative approach, we propose representing items not by their IDs but by their multi-modality information, a concept we define as Multi-Modality based Sequential Recommendation (\textbf{MMSR}). 

Recent advancements in the field have seen a shift towards directly utilizing modality representations to extract item characteristics and understand user behaviors more effectively. Innovations such as ZeroRec~\cite{ding2021zero}, UniSRec~\cite{hou2022towards}, and VQRec~\cite{hou2022learning} have been at the forefront of replacing ID representations with text representations, facilitating transferable sequential recommendations. Additionally, TransRec~\cite{wang2022transrec} and MSM4SR~\cite{zhang2023multimodal} have incorporated multi-modal information to achieve universal representations for both items and users. MMSR distinguishes itself from ID-based Sequential Recommendation (IDSR) through its inherent transferability. By utilizing the multi-modal contents of items, MMSR can seamlessly process items from various domains and platforms without relying on shareable item IDs. Moreover, MMSR addresses the cold start issue effectively by leveraging the text and visual content of items, which provides a richer semantic layer that remains untapped in strictly ID-based interactions.

We recap the basic idea of MMSR methods, which is to extract the multi-modal item embeddings from the pre-trained multi-modal item encoders (an item encoder usually consists of a visual encoder, a textual encoder, and a multi-modal fusion module) and then adopt a sequence encoder (e.g., SASRec~\cite{kang2018self}) to process the multi-modal item embeddings for recommendation training. On top of this backbone, these works design additional techniques to achieve better recommendation performance, such as enhancing the quality of text representations via parametric whitening and mixture-of-experts enhanced adaptor \cite{hou2022learning}, and achieving embedding alignment via contrastive learning with hard negatives~\cite{hou2022towards}. Though effective, these techniques are not generalized enough to fit different types of multi-modal item encoders and sequence encoders and usually require careful tuning and selecting of hyperparameters. Meanwhile, the technical breakthroughs in NLP, CV, and RecSys are always evolving, with the new state-of-the-art text encoders~\cite{devlin2018bert,liu2019roberta, ELECTRA}, vision encoders~\cite{he2016deep,huang2020pixel,he2022masked}, and sequence encoders~\cite{hidasi2015session,yuan2019simple,kang2018self,sun2019bert4rec} emerging frequently. Excessive pursuit of the additional techniques leads to poor compatibility with different types of multi-modal item encoders and sequence encoders, making it difficult to utilize the technical breakthroughs in the NLP, CV, and RecSys communities.  

To address this issue, this paper builds a simple and universal MMSR framework, which can support various types of multi-modal item encoders and sequence encoders and shows performance better or on par with the state-of-the-art multi-modal sequential recommenders. The core idea is to design a plug-and-play MMSR backbone consisting of multi-modal item encoders, multi-modal fusion modules, and the sequence encoder, such that different types of components can be seamlessly integrated into or replaced from the backbone. In our empirical study of MMSR, we try to answer the following key questions: \textbf{Q1}: How can we intuitively build MMSR from scratch, ensuring its performance on par with or even exceeds contemporary SR methods, all without the crutch of complex tricks? \textbf{Q2}: How can we build an MMSR framework such that it can directly benefit from the advances in vision-and-language communities? \textbf{Q3}: Will MMSR be supportive in addressing critical issues in the RecSys community, e.g., cold-start problems and cross-domain transferring? To answer these questions, we conduct comprehensive empirical studies along multiple dimensions, including text encoders (TE), vision encoders (VE), fusion strategies, multi-modal encoders paradigms, and RS architectures.  

To answer \textbf{Q1}, we conduct experiments on \textbf{four} large-scale industrial datasets under \textbf{three} SR architectures (i.e., SASRec~\cite{kang2018self}, GRURec~\cite{hidasi2015session}, and NextItNet~\cite{yuan2019simple}), and extensively test the performance of MMSR by equipping it with \textbf{three} TE, \textbf{three} VE and \textbf{ten} fusion strategies (including 4 vanilla coarse-grained, 4 enhanced coarse-grained and 2 fine-grained approaches). To answer \textbf{Q2}, we explore whether MMSR benefits from advances in existing Vision-Language Pre-trained (VLP) models by experimenting with different VLP architectures (e.g., CLIP~\cite{radford2021learning} \& GroupViT~\cite{xu2022groupvit},  ViLT~\cite{kim2021vilt} \& VisualBERT~\cite{li2019visualbert}, and  FLAVA~\cite{singh2022flava}) as well as various representations of vision in VLP (i.e., region-based~\cite{Faster_RCNN}, grid-based~\cite{he2016deep} and patch-based~\cite{radford2021learning}). To answer \textbf{Q3}, we put IDSR and MMSR in the cold-item scenario to evaluate their item representation ability and conduct transferring experiments on \textbf{two} source domains and \textbf{two} target domains, with the source and target datasets being scaled. In addition to exploring the above issues, practical guidance on how to effectively train MMSR would also be provided, e.g., whether full-finetuning the item encoders is necessary, what the impact of information volume brings (i.e., title length of sentences and mask ratio of images), and hyperparameter of different modules.

Our contributions can be outlined as follows:
\begin{itemize}[leftmargin=8pt, topsep = 1pt]

    \item To the best of our knowledge, we are the first to systematically explore multi-modal SR. Our observations not only thoroughly verify the effectiveness and universality of the proposed MMSR framework, but also provide much insightful guidance on developing more ID-agnostic recommendation models.

    \item We conduct comprehensive empirical studies for MMSR along multiple dimensions, including text encoders (TE), vision encoders (VE), fusion strategies, multi-modal encoders paradigms, and RS architectures, which extensively validated the reliability of our results and provide diverse insights for this domain.

    \item We bridge the divide between sequential recommendation and multi-modal learning communities, demonstrating how SR can leverage advancements in multi-modal learning to enhance recommendation quality.

    \item We evaluate the performance of MMSR using datasets from different domains and platforms. Extensive experiments demonstrate its broad applicability for cold start and transferrable SR.

\end{itemize}

\section{Related Works}

\subsection{ID-based Sequential Recommendation}

In the domain of recommender systems, ID-based Sequential Recommendation (IDSR) stands as the dominant paradigm. This approach relies on encoding unique identifiers for items and users via one-hot embeddings, serving as the foundation for numerous recommendation models. Seminal contributions in this area include works such as \cite{linden2003amazon, koren2009matrix, rendle2010factorization, he2017neural, hidasi2015session}, which have substantially influenced the development of IDSR methodologies. Our discussion further extends to state-of-the-art sequential recommenders, encompassing RNN-based models like GRU4Rec~\cite{hidasi2015session}, CNN-based models such as NextItNet~\cite{yuan2019simple}, and Transformer-based frameworks, notably SASRec~\cite{kang2018self} and BERT4Rec~\cite{sun2019bert4rec}. 

It is important to highlight that the majority of these SR approaches, albeit innovative, primarily rely on ID features—ranging from item IDs to attribute IDs. This reliance inherently subjects them to the limitations of the cold-start problem and restricts their transferability across different contexts and domains.

\subsection{IDSR with Multi-modal Side Features}
\label{sec:related_work_IDwSF}
To address this limitation, several researchers have proposed multi-modality recommenders that integrate auxiliary features to enrich item and user representations. 

Incorporating multi-modal content into IDSR, FDSA~\cite{zhang2019feature} combined text with ID-embeddings using self-attention, while CARCA~\cite{rashed2022carca} employed cross-attention to enhance item features. These methods are limited because they depend on ID-based models, reducing their effectiveness across various domains and platforms. Despite potential benefits, multi-modal recommenders still focus on items, limiting their applicability elsewhere. 

\subsection{Multi-Modal Sequential Recommendation}

Addressing the limitations above, which are constrained by their reliance on unique identifiers for items, researchers introduce a paradigm shift with Multi-Modal Sequential Recommendation (MMSR). Unlike IDSR of side features, which  primarily depends on ID information treating modal features as auxiliary, MMSR emphasizes scenarios where only multi-modal information (e.g., texts and images) serves as the input for recommendations, thereby facilitating the development of transferable RecSys by leveraging knowledge transfer across domains and platforms to mitigate data sparsity issues.

The trend towards purely modality-based recommendations is rising, with literature covering scenarios of some unimodal Modality-based SR. 
For example, ZESRec \cite{ding2021zero} employs pre-extracted text embeddings as transferable item representations for cross-task knowledge transfer, while UniSRec \cite{hou2022towards} utilizes whitening techniques \cite{su2021whitening} to bridge the gap between item text and embeddings, deriving item representations through code embedding tables. VQRec \cite{hou2022learning} matches item text with discrete index vectors for representation. MoRec \cite{yuan2023go} showing that purely modality-based models can perform comparably to ID-based models by substituting ID embeddings with text or visual representations, demonstrating improvements in feature extractors for text and image-based SR models in various scenarios. \cite{fu2023exploring} further optimized MoRec, enhancing efficiency with Adapters \cite{rebuffi2017learning}, reducing computational costs, and proving comparable performance to end-to-end methods.

Multi-modal recommender systems have been developing. TransRec \cite{wang2022transrec} optimizes hybrid modal sequences for better item representation. MISSRec \cite{wang2023missrec} incorporates user-adaptive fusion for dynamic attention to modalities. MSM4SR \cite{zhang2023multimodal} merges item images and textual keywords with visual features via BERT for user behavior analysis. \cite{peng2023multi} leverages texts, images, and prices with a re-learning strategy for task-specific challenges. \cite{ji2023online} enhances feature interactions using an ID-aware Multi-modal Transformer and online distillation. PMMRec \cite{li2023multi} employs cross-modal contrastive learning for integrating user behavior into item representations, supporting diverse learning modes.

Our MMSR framework aligns with recent advancements yet addresses the challenge of integrating diverse multi-modal item and sequence encoders. We aim to create a flexible, universally compatible, and efficient MMSR framework that works seamlessly with NLP, CV, and RecSys technologies. It has shown to surpass traditional ID-based systems in tests, proving its effectiveness in solving recommender system issues like cold starts and transfer learning.

A work closely related to ours is MMRec \cite{zhou2023comprehensive}, which outlines a clear process for multi-modal recommendation systems. However, it primarily targets GNN-based and general recommender systems, whereas our focus is on multi-modal sequential recommendation systems.

\section{Framework}
\label{Framework}
\begin{figure*}[htbp]
	\begin{center}
		\includegraphics[width=1.0\linewidth]{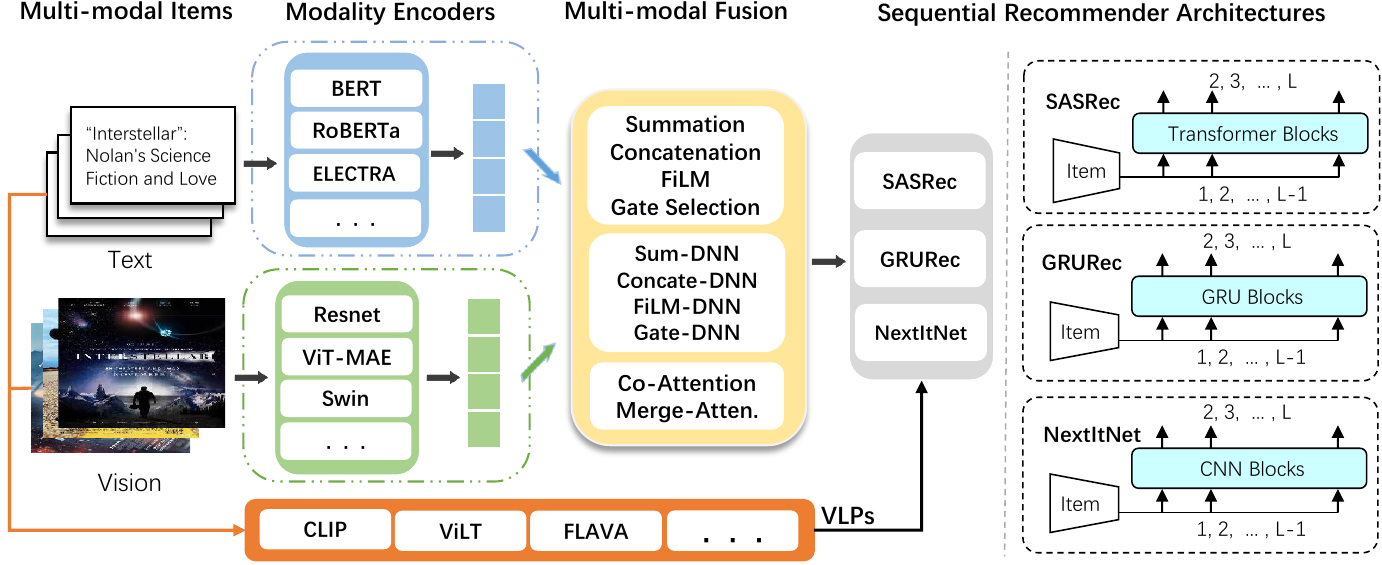}
	\end{center}
	\caption{\textbf{Illustration of MMSR}. We systematically investigate MMSR along multiple dimensions, including vision and text encoders, multi-modality fusion, multi-modal pre-training paradigms, and downstream sequential architectures.}
	\label{overview}
\end{figure*}

 In this section, we introduce essential modules in the MMSR framework and illustrate them in Figure \ref{overview}.
 For an item with textual and visual information in SR, we fine-tune the corresponding pre-trained encoders to learn its text and vision features, followed by a fusion module to combine the two unimodal features into a cross-modal representation. Note, the combination of both text and vision encoders and fusion modules can be replaced by the existing multi-modal Vison-and-Language Pre-trained (VLP) models. Finally, we feed the item representation into a sequential recommendation network to complete the downstream recommendation task.

 \subsection{Text Encoder (TE)}
 
 In this paper, we take BERT~\cite{devlin2018bert}, RoBERTa~\cite{liu2019roberta} and ELECTRA~\cite{ELECTRA} as the TE. There are two ways of using text embedding, i.e., either providing it to multiple text-specific transformers before the fusion or directly feeding it to a multi-modal fusion module. We adopt the former as our default setting, the latter will be discussed while introducing VLPs in Section \ref{VLPs_Architectures}.

\subsection{Vision Encoder (VE)}
Existing visual representations can be roughly divided into three categories:
(1) \textbf{Region Feature}. Many traditional VLP models dominantly utilize region features, which are also known as bottom-up features. They are obtained from an off-the-shelf object detector such as Faster R-CNN~\cite{Faster_RCNN}. (2) \textbf{Grid Feature}. Generated by some CNN-based methods, e.g., ResNet50~\cite{he2016deep}, it avoids very slow region selection operations.
(3) \textbf{Patch or Patch-ViT}. It is a commonly used approach that was first introduced by ViT~\cite{dosovitskiy2020image}, showing outstanding performance in image classification tasks.  In this paper, we first explore ResNet50~\cite{he2016deep}, MAE~\cite{he2022masked}, and Swin~\cite{liu2021swin} as VE in MMSR without VLPs, and then we give a detailed investigation of the three representations in VLPs in Section~\ref{various_vision_representation}. 

\subsection{Multi-modal Fusion}

As a key module that brings considerable influence in multi-modal architectures, fusion is also extensively studied in our MMSR. The fusion approaches we introduce can be divided into three classes: (1): Vanilla coarse-grained fusion methods including gate selection~\cite{gate}, concatenation, summation, and FiLM~\cite{FiLM}. (2): Enhanced coarse-grained fusion methods via adding three-layer DNN to the vanilla ones, we make these modifications given that the original fusion methods may be too shallow for large text and vision encoders. (3): Fine-grained attention-based fusion approaches, including merged attention~\cite{singh2022flava,merge_Empirical} and co-attention~\cite{Co_Decoupling, LXMERT}. For the merged attention, linguistic and visual features are simply connected and fed to $1$ (or $n$) single-stream transformer block(s). In the co-attention module, two features are fed independently into different transformer blocks, followed by a merging block to force cross-modal interaction. 

\subsection{Vision-and-language Models (VLPs)}

In this paper, SR is treated as a new downstream task of multi-modal communities. Here, we wonder if the existing VLPs paradigm can improve the task, the feasibility of which provides a theoretical potential for training specific large-scale multi-modal PTM for recommender systems. 

The multi-modal community has developed some mature VLP models to get cross-modal semantics of text and vision information.
We transfer these VLP paradigms to our MMSR to see their effect on recommendation tasks. Inspired by \cite{kim2021vilt}, VLPs can be split based on two points: (1) whether the two modalities have specific encoders; (2) whether the two modalities interact with each other in a deep fusion module. We first explore the impact of existing VLP architectures by directly introducing them into our framework, e.g., CLIP~\cite{radford2021learning} \& GroupViT~\cite{xu2022groupvit}, ViLT~\cite{kim2021vilt} \& VisualBERT~\cite{li2019visualbert} and FLAVA~\cite{singh2022flava} (They stand for three typical paradigms of VLPs, i.e., two-stream and shallow fusion, single stream and deep fusion, two-stream and deep fusion). 

In our experiments, we note that variations in the architecture of vision models significantly affect the overall performances. Therefore, we carry out further exploration by evolving vision representation in VLPs. We select VisualBERT~\cite{li2019visualbert} and replace the vision input with region-based (Faster-RCNN~\cite{Faster_RCNN}), grid-based (ResNet50~\cite{he2016deep}), and patch-based (CLIP-ViT~\cite{radford2021learning}) visual representations in Section~\ref{various_vision_representation}.

\subsection{Recommendation Architectures}

To evaluate the MMSR framework, we systematically selected three representative SR architectures: Transformer-based SASRec~\cite{kang2018self}, GRU-based GRURec~\cite{li2022recguru}, and CNN-based NextItNet~\cite{yuan2019simple}. We analyze the synergy between these popular sequential recommendation architectures and multi-modal item encoders, focusing on their ability to integrate and leverage multi-modal data effectively. We use joint distributions to capture comprehensive sequence relationships, as shown in the equation: \begin{equation}
           R(x)  = \prod \limits_{i=0}^l r(x_{i}| x_{0:l-1}, \theta)r(x_{0})
           \tag{3}
\end{equation}
where, $r(x_{i}|x_{0:l-1}, \theta)$ represents the conditional probability of the $i$-th item, given all preceding items $x_{0:i-1}$.


We experiment with various hyperparameters: the number of layers is set to 2, 4, and 2 for SASRec, GRURec, and NextItNet, respectively, after experimentation with options in $\{1,2,4,8\}$. For SASRec, we configure the attention heads to 2, aiming to optimize the interaction between sequential processing and multi-modal integration for enhanced recommendation performance.

\section{Experiments}
\subsection{Datasets}
Many existing modalitiey-related SR \cite{hou2022towards,hou2022learning} primarily focus on the e-commerce domain of item images with clean and pure backgrounds (e.g., items in the Amazon dataset all belong to the "fashion" category). However, There is insufficient evidence on whether complex multimodal information can facilitate SR with complex  (e.g. short videos with broad topics).content
To provide a more comprehensive evaluation, we evaluate MMSR on \textbf{four} real-world datasets, i.e., the \textbf{HM}\footnote{https://www.kaggle.com/competitions/h-and-m-personalized-fashion-recommendations/overview} and \textbf{Amazon}\footnote{https://nijianmo.github.io/amazon/index.html} datasets for clothing purchase, the \textbf{Bili}\footnote{https://www.bilibili.com/} and \textbf{Kwai}\footnote{https://www.Kwai.com/?isHome=1} dataset from short-video platforms. For HM, we use descriptions and covers of the products and add categorical tags to the textual features to distinguish products with the same descriptions. For Bili and Kwai datasets, the title and cover of each video are regarded as the modality input. We consider those with less than 10 occurrences in the training set to select cold items in the cold scenarios. 

The HM and Amazon datasets include customer-product (user-item) interactions and customer and product metadata, e.g., category and color, descriptions, and covers. In this paper, we only use texts and images, adding some category information as part of the description to further distinguish different items. Following \cite{yuan2019simple}, we removed users with less than 5 interactions simply, and randomly selected 200K and 100K users.  

To build the Bili dataset, we randomly crawl videos (less than 10 minutes in duration) of different categories from the online platform. We record the public comment information of these videos as interactions (i.e., we think the user would be interested in this video if a user comments on a video). We do not capture any user's personal information about privacy, whose IDs are already anonymous after data masking. The video metadata such as cover and title are recorded. The Kwai dataset is collected similarly to the Bili dataset. 
Then we also removed users with less than 5 interactions and randomly selected about 100K and 200K users from the Bili and Kwai datasets, respectively. 
 
We take some basic pre-processing by setting the size of all images to 224 × 224, with a maximum of 50 words for all text descriptions or titles (covering more than 95\% of the descriptions).  Noted, the datasets used in this paper should ensure that the user's decision (i.e., whether to interact with the item or not is determined only by the modality characteristics of the item). But in fact, none of the 2 datasets meets this assumption. For example, except for the textual descriptions and cover of the products, users' purchase decisions on the HM dataset are also influenced by the price of the item. However, we can still achieve better performance even taking only text and image representations of items, which proves the potential of MMSR over IDSR.

Besides, for all datasets, we consider cold scenarios (different from regular scenarios). That is, we count the interactions of all items in the training set and consider those items with less than 10 occurrences as cold items~\cite{yuan2023go}.  Then, we truncate the complete user sequence to get some sub-sequences with a cold item at the end, all of which are used to evaluate the capability of MMSR in the cold scenario. The preprocessed dataset statistics are presented in Table ~\ref{Statistic_datasets}.
\begin{table}
\caption{Dataset statistics (after preprocessing). Specifically, \textbf{users}, \textbf{items}, and \textbf{cold} denote the numbers of users, items, and truncated users sub-sequences with a cold item at the end, respectively. The \textbf{avg.} denotes the average sequence length of users' historical interactions.}
\begin{center}

\begin{tabular}{lccccc}
\toprule
       Dataset & \#users & \#items & \#cold & avg.  & types \\
\midrule
       HM & 200K & 85,019 & 204,988 & 31.39 & purchases\\
       Amazon    & 100K   & 63,456 &  42,674 &  7.42 & purchases\\
       Bili & 100K & 76,178 & 5,632 & 17.95 & comments \\
       Kwai & 200K & 39,410 & 83,689 & 13.45 & comments \\
\bottomrule
\end{tabular}

\label{Statistic_datasets}
\end{center}
\end{table}
\label{Datasets}

\subsection{Default Implementation Details}
\label{Implementations}

Our principle is to ensure that MMSR is always thoroughly tuned, including the learning rate $\in$ \{1e-4, 8e-5, 5e-5, 3e-5\}, dropout rate $\in$ \{0.1, 0.3, 0.5\}, and batch size $\in$ \{64, 128, 256\} and fixed embedding dimension of 768. 
All transformer-based text and vision encoders are fine-tuned with only the top two transformer blocks (three CNN blocks for ResNet50~\cite{he2016deep}) to ensure the trainable parameters of similar magnitude. Note that full fine-tuning does not guarantee better results and is even worse (see Section \ref{bag_of_tricks}). 

We train MMSR with AdamW optimizer for 150 epochs. The learning rate decays linearly to 10$\%$ of the original learning rate after 120 epochs. The weight decay is searched in \{0, 0.001, 0.01, 0.1\} and finally set to 0 for item encoders and 0.1 for the rest of MMSR. All experiments are carried out on a single server with  NVIDIA A100 GPUs. Without specifying, the reported MMSR may not achieve its best possible performance because it is very expensive and time-consuming to search for all possible hyperparameters. 

Following \cite{krichene2022sampled}, we rank the ground-truth target item by comparing it with all the other items in the item pool. Finally, we report the results of the testing set by choosing checkpoints with the best validation performance.

\subsection{Baselines}
To provide a comprehensive evaluation on the performance of our method, we compare MMSR with the following baseline methods from three different groups\footnote{We note that previous works seldom consider the multi-modal setting involving both texts and images. For a fair comparison, we select the strongest baselines (CARCA from group 2 and MoRec from group 3), and improve them into their multi-modal versions (denoted as CARCA++ and MoRec++).}:

\noindent\textbf{Pure ID-based sequential recommenders (IDSR)} assign a unique ID to each item in the dataset. 
Item IDs are subsequently converted into embedding vectors and then fed into a sequence encoder to extract user preferences. However, as item IDs are not shareable across domains and platforms, these methods are difficult to transfer to new recommendation scenarios. 
Typical examples are listed as follows:
\begin{itemize}[leftmargin = 8pt, topsep = 1pt]
\item\textbf{GRURec} \cite{hidasi2015session} employs RNN as the sequence encoder and introduces ranking loss functions for model optimization.
\item\textbf{NextItNet} \cite{yuan2019simple} is a CNN-based sequential recommender, which combines masked filters with dilated convolutions to increase the receptive fields.
\item\textbf{SASRec}~\cite{kang2018self} utilizes a unidirectional Transformer as the sequence encoder, which flexibly assigns attention weights to different items in the sequence.
\end{itemize}
\noindent \textbf{ID-based sequential recommenders with side features (IDSR w. side feat.)} enhance pure ID-based sequential recommenders with multi-modal side features such as texts and images. 
The core idea is to integrate multi-modality contents into the backbone of ID-based sequence encoders in order to improve recommendation performance. 
Such designing is still constrained under the ID-based paradigm, thus still falls short of transferability. 
Typical examples are listed as follows:
\begin{itemize}[leftmargin = 8pt,topsep = 1pt]

\item\textbf{FDSA} \cite{zhang2019feature} captures the transition patterns between items and the contextual features of items (e.g., brands, text descriptions) via feature-based self-attention blocks.
\item\textbf{CARCA++} is our improved version of CARCA \cite{rashed2022carca}. 
It utilizes cross-attention to capture the correlation between item IDs and contextual multi-modal features. 
The original version only supports imaged-based features, and we improve it into a multi-modal version.
\end{itemize}

\noindent\textbf{Modality-based sequential recommenders} learn transferable item representations via item texts or images. 
Typical examples are listed as follows:
\begin{itemize}[leftmargin = 8pt, topsep = 1pt]
\item\textbf{VQRec} \cite{hou2022learning} maps item texts into discrete codes and then performs the lookup operation in the code embedding table to learn transferable item representations.
\item\textbf{MoRec++} is our improved version of MoRec \cite{yuan2023go}. 
For a fair comparison, we improve it into a multi-modal version by fusing the representations extracted from text and the vision encoders.
\end{itemize}

\subsection{Loss and Negative Sampling }
We adopt the same negative sampling strategy  \footnote{Some literature \cite{yuan2016lambdafm,chen2022cache} shows other advanced negative sampling strategies, which does not affect the fairness of our work.} in \cite{yuan2023go} which is absolutely fair to MMSR and other SR methods. 
\textbf{For consistency in our evaluation, we employ the same loss function and sampling technique for MMSR and above baseline methods}. Specifically, a negative sample is randomly selected \cite{yuan2023go} and binary cross-entropy function as the loss. For each observed interaction $<u, j>$,  it can be formulated as follows:

\begin{equation}
		min
		- \sum_{<u, i, j> \in \mathcal{S}^{sample}}
		\left\{
		\log(\sigma(\hat{y}_{ui}))
		+
		\log(1- \sigma(\hat{y}_{uj}))\right\},
   \tag{6}
	\end{equation}
 where $\sigma(x) = 1/(1+e^{-x})$.

 \subsection{Evaluation Metrics}
\label{Metrics}
We adopt the standard leave-one-out strategy \cite{kang2018self}
and split the three used datasets into three parts: training, validation, and testing data, and evaluate all models in terms of two Top-N ranking metrics: Hit Ratio (HR@10) and Normalized Discounted Cumulative Gain (NDCG@10, or NG@10 for short).



\section{Performance of MMSR (Q1)}

To systematically explore how to build a general MMSR framework to achieve the best results, we conduct extensive experiments using (1) different text and vision encoders, (2) different multimodal fusion methods, and (3) different SR architectures. The best MMSR architecture is then comprehensively compared with current baselines.

\subsection{Explorations of Various Item Encoders}   
\label{Explorations_various_encoders}
To explore the potential of MMSR step by step, we first try different combinations of text and vision encoders without applying existing VLP models. Concretely, we explore the potential of our MMSR in unimodal mode\footnote{Note that MMSR-T and MMSR-V in the tables are the ablation versions of our MMSR with only the text and vision information.}, then we construct the complete multi-modal MMSR by merging both TE and VE, i.e., we set RoBERTa~\cite{liu2019roberta} as the default TE and combine it with various VE, and choose MAE~\cite{he2022masked} as the default VE to integrate with other TE in reverse. For all of them, we select a one-block merge attention module as the fusion approach. All explorations are evaluated over HM and Bili datasets with SASRec, and the results are reported in Table~\ref{tb:various_text_mae_recommendation}.

\begin{table}[htbp]
   \caption{Comparisons of MMSR with different TE(i.e., BERT~\cite{devlin2018bert}, RoBERTa~\cite{liu2019roberta} and ELECTRA~\cite{ELECTRA}) and VE (i.e., ResNet50~\cite{he2016deep}, MAE B-224/16~\cite{he2022masked},  and Swin B-224/16~\cite{liu2021swin}). All TE are in the base size. RoBERTa and MAE are default TE and VE, respectively. }
   \begin{center}
  \setlength{\tabcolsep}{3.5pt}{
   \begin{tabular}{lccccc}
   \toprule
   \multirow{2}{*}{-} & \multirow{2}{*}{Encoders} &\multicolumn{2}{c}{HM }  &\multicolumn{2}{c}{Bili}\\
   \cmidrule(lr){3-4}\cmidrule(lr){5-6}
    &  &HR@10 &NG@10 &HR@10 &NG@10 \\ 
   \midrule
   \multirow{3}{*}{MMSR-T}& RoBERTa &8.2275 &4.6923 &3.7790	&1.9655 \\
   & ELECTRA &8.2010 &4.6195 & 3.3690 &1.7178\\
   &BERT &8.1810 &4.6353 &3.7990 &1.9810\\
   \midrule

   \multirow{3}{*}{MMSR-V}& ResNet	&7.3520 &3.9049 &3.2010 &1.6550 \\
   & MAE &6.1945 &3.2329 &2.4200 &1.1823 \\
   &Swin &7.8615  &4.2045 &3.3940 &1.7281 \\
   \midrule

    \multirow{6}{*}{MMSR}& Ro.+ResNet &8.4280	&4.6649 &3.6600 &1.8942 \\
   & Ro.+Swin	&\textbf{10.1095} &  5.7703 &   4.0810 & \textbf{2.2322} \\
   & Ro.+MAE &9.9635 &\textbf{5.8046} &\textbf{4.2090} &2.1194 \\
   
   & ELE.+MAE &9.1675 &5.1984 &3.7940 &1.9392 \\
   & BERT+MAE &9.1100 &5.1590 &4.1300 &2.1484 \\
   \bottomrule

\end{tabular}}

   \label{tb:various_text_mae_recommendation}
\end{center}
\end{table}

\textbf{Various Text Encoders.} We do not observe significant differences in performance for different TE in both MMSR-T and MMSR. 
This suggests that the capabilities of various TEs in feature representation have become very close, indicating that text representation techniques have matured considerably.

\textbf{Various Vision Encoders.} MAE~\cite{he2022masked} or Swin~\cite{liu2021swin} perform the best in MMSR. However, the results of MAE~\cite{he2022masked} and ResNet~\cite{he2016deep} \footnote{Like PixelBERT~\cite{huang2020pixel}, ResNet is used as the backbone to reduce the dimension of pixel features and we selected 49 vectors from each image.} are exactly opposite. For the poor performance of ResNet~\cite{he2016deep} in MMSR, we guess it is because the 
current fusion approach cannot digest the outputs of ResNet~\cite{he2016deep} (CNN-based) and RoBERTa~\cite{liu2019roberta} (transformer-based) well. For MAE~\cite{he2022masked}, the default high mask ratio (75\%) may be the reason for its poor performance in MMSR-V.

\textbf{Results.}  The performance of MMSR-T is generally better than that of MMSR-V, which may be that the dense semantics of textual information is easier to learn in SR~\cite{hou2022towards}. The fully transformer-based (e.g., RoBERTa and MAE or Swin) perform the best in MMSR, which suggests the superiority of unified architectures. The MMSR with a comprehensive combination of text and vision encoders demonstrates its superiority.


\begin{table}[htbp]
   \caption{Comparison of MMSR with various fusion methods. RoBERTa (base) \cite{liu2019roberta} and MAE B-224/14 \cite{he2022masked}) are used as text and vision encoders, respectively. 
   \textbf{Coarse}: MMSR with coarse fusion strategies. \textbf{Coarse+}: add a three-layer DNN based on the coarse fusion. \textbf{CoAtten}-1: co-attention fusion of 1 layer. \textbf{MergeAttn}-$n$: merge-attention fusion of $n$ layer. }
   \begin{center}
\setlength{\tabcolsep}{3pt}{\begin{tabular}{lccccc}
   \toprule
   \multirow{2}{*}{-} & \multirow{2}{*}{Fusion} &\multicolumn{2}{c}{HM}  &\multicolumn{2}{c}{Bili}\\
   \cmidrule(lr){3-4}\cmidrule(lr){5-6}
    &   &HR@10 &NG@10 &HR@10 &NG@10 \\ 
   \midrule
   \multirow{4}{*}{Coarse} &Gate &8.9505   &5.0762 &3.9120  &2.0108 \\
   & FiLM	&8.6250   &4.8787 &3.8300  &1.9809 \\
   & Summation &6.9605 &3.7458 &3.4780 &1.7709  \\
   & Concatenation &6.7575  &3.6387 &3.4060  &1.7318 \\
   \midrule
   \multirow{4}{*}{Coarse+} &Gate &8.4120 &4.7515 &3.5300 &1.7819 \\
   &FiLM &8.4760  &4.7998 &3.5340  &1.7711 \\
   &Summation &8.5270 &4.8214 &3.5960 &1.8021 \\
   &Concatenation &8.5035 &4.8302 &3.3690 &1.6861\\
   \midrule
   \multirow{4}{*}{Attention}& CoAttn-1 &7.6810 &4.1932 &3.8210  &1.9463 \\
   & MergeAttn-1 &9.9635 &\textbf{5.7703} &\textbf{4.0810} &\textbf{2.1194} \\
   &MergeAttn-2 &\textbf{10.0780}  &5.7448 &3.8440 &1.9844 \\
   &MergeAttn-4 &9.8975 &5.7634 &3.4210 &1.6852 \\
   \bottomrule
\end{tabular}}
   \label{tb:fusion}
\end{center}
\end{table}

\begin{table}[htbp]
   \caption{Comparison of IDSR and MMSR with various SR architectures. HR@10 is as the default metric here. RoBERTa (base)\cite{liu2019roberta} and Swin B-224/14 \cite{liu2021swin}) are selected as the text and vision encoders. The best results are bolded. ‘Improv.’ is the relative improvement of the MMSR compared to the best IDSR.
   }
   \begin{center}
    \setlength{\tabcolsep}{1.5pt}{
   \begin{tabular}{llcccc}
   \toprule
   \multirow{2}{*}{Dataset} & \multirow{2}{*}{Architecture} & \multirow{2}{*}{IDSR} &\multicolumn{3}{c}{MMSR}   \\
   \cmidrule(lr){4-6} 
   & 
    & &MMSR-T &MMSR-V &MMSR \\
   \midrule

   \multirow{3}{*}{Bili}
   &SASRec
   &  3.3942 &3.7790 &3.3640 &\textbf{4.2090}  \\
   &GRURec 
   &  2.3247  &2.9260  &2.6400  &\textbf{3.2960}  \\
   &NextItNet 
   & 2.0521 &2.3860  &1.8920  &\textbf{2.7120}  \\
   \midrule

   \multirow{3}{*}{Kwai}
   &SASRec
   &4.7521 &5.3110&4.7485 &\textbf{5.5260} \\
   &GRURec 
   & 3.9487	& 4.2845  & 4.1315  &\textbf{4.7190} \\
   &NextItNet
   & 3.1538	&  3.7525  & 3.4195  &\textbf{4.6025} \\

    \midrule
   \multirow{3}{*}{HM}
   &SASRec
   & 8.4672 &8.2670 &7.8615 &\textbf{10.1095} \\
   &GRURec 
   & 6.1241 & 5.2780   &5.7215 &\textbf{6.7965} \\
   &NextItNet 
   &  6.1752  &6.5275  &  5.8635 & \textbf{7.7755}  \\

   \midrule
      \multirow{3}{*}{Amazon}
   &SASRec
   & 19.1250 &19.2436 &19.1529 &\textbf{20.1095}  \\
   &GRURec 
   &  16.0417& 16.0056 &15.0926 &\textbf{18.3956}  \\
   &NextItNet 
   &  15.2378  &16.2342  &  15.9076 & \textbf{17.7984} \\
   \bottomrule
   \end{tabular}}
   \label{tb: various_Rs}
   \end{center}
\end{table}

\begin{table*}[!htbp]
\caption{Performance comparison (\%) on four datasets. The best and the second-best performance in each row are bolded and underlined, respectively. Improvements compared with the best baseline method are indicated in the last column.}
\label{various_representation}
\begin{center}
  \setlength{\tabcolsep}{5pt}{
    \begin{tabular}{llccccccccc}
   \toprule
   \multirow{2}{*}{Dataset} & \multirow{2}{*}{Metrics} &\multicolumn{3}{c}{IDSR} & \multicolumn{2}{c}{IDSR w. Side Feat.} & \multicolumn{2}{c}{Modality-based SR}  &\multicolumn{1}{c}{Ours} &\multirow{2}{*}{Improv.} \\
   \cmidrule(lr){3-5} \cmidrule(lr){6-7} \cmidrule(lr){8-9}  \cmidrule(lr){10-10}
   &  &GRURec &NextItNet & SASRec & FDSA  & CARCA++   & VQRec   &MoRec++ & MMSR \\ 
\midrule
    \multirow{6}{*}{Bili}
    &HR@10   & 2.3247 & 2.0521 & 3.3942 &  3.7623  &  \underline{4.0456}   & 1.3569
            & 3.7531 & \textbf{4.2210}    & +4.16\%\\
    &HR@20   & 3.3678 & 3.1256 & 4.4350 &  4.9143  &  \underline{6.1476} & 2.0721
            & 5.8824 & \textbf{6.3135}    & +2.63\%\\
    &HR@50   & 6.0234 & 5.1288 & 7.9087 & 8.9260  &  \underline{9.9689}  & 3.5731
            & 8.8856 & \textbf{10.4437}    & +5.29\%\\
            
    &NDCG@10 & 1.2145 & 1.0378 & 1.6729 & 1.7934  &  \underline{2.1159}  &  0.6089
            & 1.9834 & \textbf{2.2315} & +5.17\%\\
    &NDCG@20 & 1.4271 & 1.2876 & 2.0174 & 2.2837  &  \underline{2.5765}   &  0.7167
            & 2.4829 & \textbf{2.7046} &  +4.74\%\\
    &NDCG@50 & 1.9463 & 1.8291 & 2.6175 &  2.9750  &  \underline{3.3290}  & 1.0827
            & 3.2254 & \textbf{3.5287}    & +5.66\%\\

    \midrule
    \multirow{6}{*}{Kwai}
    &HR@10  &3.9487 & 3.1538 & 4.7521 & 4.9357 & \underline{4.9571}  & 1.9523 & 4.9518 &\textbf{5.3786} & +7.84\%\\
    &HR@20 & 5.9744 & 5.1966 & 6.8462 & 6.9929 & \underline{7.1286} & 2.9568 & 6.7517 &\textbf{7.7794} & +8.37\%\\
    &HR@50 & 9.8718 & 8.6068 & 10.9051 & 10.9714 & 11.3502  & 4.8528 & \underline{11.7659} &\textbf{12.3295} & +4.57\%\\
    &NDCG@10 &2.0598 & 1.9915 & 2.5043 & 2.6664 & 2.5857  & 0.8732 & \underline{2.6297} & \textbf{2.8576} & +7.99\%\\
    &NDCG@20 & 2.5641 & 2.2137 & 3.0513 & 2.8317 & 3.0857  & 1.2083 & \underline{3.1294} & \textbf{3.4085} & +8.19\%\\
    &NDCG@50 & 3.3419 & 2.9744 & 3.8556 & 3.8956 & \underline{4.0652}  & 1.5093 & 4.0086 & \textbf{4.3092} & +5.66\%\\

    \midrule
    \multirow{6}{*}{HM}
    &HR@10  & 6.1241 & 6.1752 & 8.4672 & 8.8214 & \underline{9.7678} & 4.1623 & 9.6933 & \textbf{10.0425} & +2.82\%\\
    &HR@20 & 8.3577 & 8.3066 & 11.0511 & 11.2213 & \underline{12.1035}  & 5.7416 & 11.9867 & \textbf{12.6694} & +4.47\%\\
    &HR@50 & 12.6058 & 12.7372 & 15.5547 & 15.4623 & \underline{16.6235}  & 8.0028 & 16.0933 & \textbf{17.5643} & +5.36\%\\
    &NDCG@10 & 3.6350 & 3.5328 & 5.4715 & 5.6917 & \textbf{6.4234}  & 2.2235 & 6.1400 & \underline{6.3646} & -0.91\%\\
    &NDCG@20 & 4.1898 & 4.0431 & 6.1153 & 6.4784 & \underline{6.8217}  & 2.6315 & 7.0533 & \textbf{6.9224} & +1.46\%\\
    &NDCG@50 & 5.0292 & 5.0146 & 7.0146 & 6.6347 & \underline{7.5158} & 3.0562 & 7.0867 & \textbf{7.8824} & +4.86\%\\

    \midrule
    \multirow{6}{*}{Amazon}
    &HR@10 & 16.0417 & 15.2378 & 19.1250 & 18.7667 & \textbf{19.7250}  & 17.7143 & 19.2512 & \underline{19.6412} & -0.42\%\\
    &HR@20 & 17.6256 & 15.9234 & 19.3078 & 18.8245 & \underline{19.8567}  & 19.0567 & 19.5923 & \textbf{19.9934} & +0.68\%\\
    &HR@50 & 19.3023 & 17.3445 & 20.9645 & 19.4423 & \underline{21.3012}  & 20.4012 & 21.2578 & \textbf{21.9023} & +2.81\%\\
    &NDCG@10 & 14.9923 & 12.9945 & 16.7067 & 15.8512 & \underline{17.1434}  & 12.8034 & 17.0767 & \textbf{17.3656} & +1.30\%\\
    &NDCG@20 & 15.2412 & 13.2323 & 16.9345 & 16.3145 & \underline{17.8312}  & 13.0434 & 17.9245 & \textbf{18.0201} & +1.20\%\\
    &NDCG@50 & 15.5745 & 13.5012 & 17.2312 & 17.0278 & \underline{18.5145}  & 13.3923 & 18.3934 & \textbf{18.6512} & +0.77\%\\

\bottomrule
\end{tabular}}
\end{center}
\end{table*}

\subsection{Impact of Various Fusion Approaches} 


In this section, we show investigations on various multi-modal fusion strategies for MMSR with a default of TE, VE and RS architecture (i.e., RoBERTa~\cite{liu2019roberta}, MAE~\cite{he2022masked}, and SASRec~\cite{kang2018self}).
Specifically, we explore four vanilla coarse-grained fusion methods (i.e., gate selection~\cite{gate}, concatenation, summation, and FiLM~\cite{FiLM}), four enhanced coarse-grained fusion methods (with a three-layer DNN applied to the previous ones), and two fine-grained fusion methods (i.e., co-attention~\cite{Co_Decoupling} and merge-attention~\cite{merge_Empirical}).

\textbf{Coarse-grained fusion.} As shown in Table \ref{tb:fusion}, MMSR with coarse-grained fusion lacks competitiveness, especially in the HM dataset. Among them, gate~\cite{gate} and  FiLM~\cite{FiLM}  do slight favor to MMSR while summation and concatenation seriously deteriorate the performance. Comparing \textit{Coarse} with \textit{Coarse++}, we notice that added DNN drives all the shallow fusions to similar performance. It seems the three-layer DNN overwhelms the multi-modal fusion.

 \textbf{Fine-grained fusion.} The results in Table \ref{tb:fusion} show that merge-attention outperforms co-attention. We also find that there is no improvement even though we deepen the merge-attention from one to four transformer blocks, and the results on the Bili dataset are even inversely proportional to the depth. It may be that the amount of data is not enough to support the learning of deep fusion. It suggests that VLP models should be considered if we want to use deeper fusion modules (see Section \ref{VLPs_Architectures}).

\textbf{Results.} Coarse-grained fusion does not guarantee the performance of MMSR. As for fine-grained methods, merge-attention exceeds co-attention in integrating different modalities.

\subsection{MMSR with Various SR Architectures}  

Based on the above explorations, we choose powerful RoBERTa~\cite{liu2019roberta} and Swin~\cite{liu2021swin} as text and vision encoders, respectively. To validate the robustness of our MMSR, we consider the effect of various architectures, including SASRec~\cite{kang2018self}, GRURec~\cite{hidasi2015session}, and NextItNet~\cite{wang2022transrec}, as shown in Figure \ref{overview} (right). We evaluate the performance over \textbf{four} recommendation datasets (i.e., HM, Bili, Kwai, and Amazon). 

 \textbf{Results.}  As shown in Table \ref{tb: various_Rs}, MMSR configured with SASRec shows strong competitiveness. Furthermore, no matter which SR architecture is given, the performance of traditional ID-based SR (IDSR) are much lower than that of MMSR. The ablation study shows that single-modal information, is still very competitive, even surpassing IDSR in some cases. This suggests that current CV and NLP techniques have enabled id-agnostic recommendation models to excel, highlighting the potential of the MMSR paradigm.

\subsection{Overall Performance Comparisons}
We evaluate MMSR and other baseline methods on 4 source datasets. From the results of Table~\ref{various_representation}, we can reach the following conclusions:
\begin{itemize}[leftmargin = 8pt, topsep = 1pt]
    \item Against other baselines, MMSR consistently comes out on top in most tests. What's more, it performs just as well, if not better than, methods that rely on IDs, even those with added side features. This suggests that MMSR can make sense of multi-modal data for recommendations without needing item IDs.
    \item Compared with the best baseline CARCA++, MMSR shows smaller performance gains on HM and Amazon datasets, but larger gains on Bili and Kwai datasets. This is probably because the former two datasets have relatively clean and pure image backgrounds, while the latter has more complex visual presentations (e.g., posters). Benefiting from the powerful and robust backbone derived from the above sufficient exploration strategy, MMSR can effectively cope with data noise and improve performance, showing greater performance improvements on the Bili and Kwai datasets.
    \item  Our experiments show that MoRec and our proposed method outperform VQRec, indicating that end-to-end learning is more effective than traditional pre-extraction. This is likely due to the fact that end-to-end learning allows the item encoder to better capture the semantic information of modalities \cite{yuan2023go}.

\end{itemize}

\section{Inherit Advances in Multi-modal Pre-training Community (Q2)}



MMSR employs SR as a subsequent task in multimodal learning, prompting us to examine if advancements in the multi-modal pre-training community, i.e., Vision-and-Language Pre-training (VLP), can enhance MMSR. In particular, we delve into two fundamental questions frequently addressed in the VLP community: (1) the architecture of VLP model \cite{singh2022flava,kim2021vilt}, and (2) the visual representation within VLP model~\cite{kim2021vilt}.

\subsection{Impact of Various VLP Architectures} 
\label{VLPs_Architectures}
\begin{table}[htbp]
   \caption{Comparison of MMSR with various VLP models of different architectures. (a): two-stream \& shallow fusion. (b): single stream \& deep fusion. (c): two-stream \& deep fusion.}
   \begin{center}
     \setlength{\tabcolsep}{3.5pt}{\begin{tabular}{lccccc}
   \toprule
   \multirow{2}{*}{} & \multirow{2}{*}{VLPs} &\multicolumn{2}{c}{HM }  &\multicolumn{2}{c}{Kwai}\\
   \cmidrule(lr){3-4}\cmidrule(lr){5-6}
    &  &HR@10 &NG@10 &HR@10 &NG@10 \\ 
   \midrule
   \multirow{2}{*}{a}& CLIP & 6.4240 &3.3109 &4.2510 &2.1332  \\
   & GroupViT	&7.7935 &4.2868 &4.1800 &2.1132 \\ 
   \midrule
   \multirow{2}{*}{b}&ViLT	&9.3465 &5.3226 &4.9540 &2.5938 \\
   & VisualBERT &9.2680	&5.2937 &5.0740 &2.6141\\
   \midrule
   \multirow{1}{*}{c}&FLAVA &\textbf{10.1435} &\textbf{6.5820} & \textbf{5.4040} &\textbf{2.7125}\\
   \bottomrule
\end{tabular}}
\label{tb:various_VLP_architecture}
\end{center}
\end{table}
Our exploration initiates with a meticulous examination of diverse VLP architectures. To be specific, these can be broadly classified into three emblematic paradigms: (1) Two-stream and shallow fusion, as exemplified by models such as CLIP~\cite{radford2021learning} and GroupViT~\cite{xu2022groupvit}, (2) Single-stream and deep fusion, represented by the likes of ViLT~\cite{kim2021vilt} and VisualBERT~\cite{li2019visualbert}, and (3) Two-stream and deep fusion, encapsulated by models like FLAVA~\cite{singh2022flava}.


Specifically, we choose SASRec as the default RS architecture. For CLIP and GroupViT, we fine-tune two transformer blocks for the text and vision encoders and use a concatenation layer to connect both as the input of the downstream recommendation network. Since VisualBERT can not extract vision features itself, we use Faster-RCNN\cite{Faster_RCNN}  with a threshold of 0.05 and select 49 regions from each image as visual input. The last 4 transformer blocks of VisualBERT are fine-tuned, which is consistent with ViLT.  As for FLAVA, we fine-tune the top two transformer blocks of text and vision encoders and all fusion modules. 

\textbf{Results.} In Table \ref{tb:various_VLP_architecture}, we can see that MMSR performs worst in the CLIP \& GroupViT architecture. By analyzing in conjunction with the other two VLPs, we hypothesize it is the shallow fusion that deteriorates the performance. We also notice that FLAVA drives MMSR to completely surpass other VLPs, which is consistent with its excellent performance in the multi-modal fields. Note, the best-performing MMSR in RQ(1) has a similar architecture to FLAVA, which may imply that the success of PTM in other domains can be transferred to the recommendation task.

\subsection{Impact of Various Vision Representation} 
\label{various_vision_representation}

We carry out further exploration by evolving vision representation in VLPs (see Section \ref{Framework} for details). We select VisualBERT~\cite{li2019visualbert} as our foundational model due to its iconic status within the domain of VLPs. Most subsequent strides of vision representation are indeed enhancements built upon it. To validate the MMSR, we have incorporated an assortment of vision representations, including region-based (Faster-RCNN~\cite{Faster_RCNN}), grid-based (ResNet50~\cite{he2016deep}), patch-Embedding (CLIP-ViT~\cite{radford2021learning}, excluding 12 transformers) and patch-Vision Transformer (CLIP-ViT, including 12 transformers). These representations typify a significant trajectory in the evolution of VLPs, thereby prompting our curiosity to investigate their adaptability to MMSR.

\textbf{Results.} As is shown in Table \ref{tb:various_vision_VLPS}, all MMSRs with various vision representations can exceed IDSR, and patch-ViT representation is the best, which is in line with the development process in the field of VLPs. Region-based and Patch-Emd representations underperform, whereas MMSR with Grid-based and Patch-ViT demonstrate notable improvements over IDSR.  This disparity might be attributed to the strong augmentation of the latter two specific vision encoders, namely ResNet50~\cite{he2016deep} and CLIP-ViT \cite{radford2021learning}. Note that in Section \ref{Explorations_various_encoders}, ResNet50 (grid-based) performs poorly in the previous setting. However, it achieves good results here, suggesting that deep fusion may resolve instability caused by inconsistent representations.

\begin{table}[htbp]
   \caption{Comparison of vision representations for MMSR. }
   \begin{center}
     \setlength{\tabcolsep}{2pt}{
   \begin{tabular}{lccccc}
   \toprule
   \multirow{2}{*}{-} & \multirow{2.1}{*}{Representations} &\multicolumn{2}{c}{HM}  &\multicolumn{2}{c}{Kwai}\\
   \cmidrule(lr){3-4}\cmidrule(lr){5-6}
    &   &HR@10 &NG@10 &HR@10 &NG@10 \\ 
   \midrule	
   \multirow{1}{*}{IDSR}& - &8.4672 	&5.4715 &4.7521 & 2.5043\\
   \midrule
   \multirow{4}{*}{VisualBERT}& Region  &9.2680	&5.2937 &5.0740 &2.6141 \\
   &Grid	&9.5875  &5.5167 &5.2250 &\textbf{2.6814} \\
   &Patch-Emb	&9.0205 &5.1664 &5.0695 &2.6255 \\
   &Patch-ViT	&\textbf{9.8525} &\textbf{5.7026} & \textbf{5.2502} &2.6374
 \\
   \bottomrule
\end{tabular}}
\label{tb:various_vision_VLPS}
\end{center}
\end{table}

\section{Cold-start and Transferability (Q3)}
\begin{table}
   \caption{Performance comparison (\%) under cold-start setting.}
   \begin{center}
   \small
\setlength{\tabcolsep}{5pt}{
   \begin{tabular}{llcccc}
   \toprule
   \multirow{1}{*}{Dataset} & \multirow{1}{*}{Metrics} & SASRec &MMSR-T &MMSR-V &MMSR \\
    \midrule

    \multirow{2}{*}{Bili}
   &HR@10
   & 0.1192  &\textbf{1.5493}  &0.9296  &1.3824\\
   &NG@10 
   & 0.0450  &\textbf{0.7722} &0.4706  & 0.7603\\
   \midrule
   
   \multirow{2}{*}{Kwai}
   &HR@10
   & 0.0419 &3.9812  &3.9408  &\textbf{4.7393}\\
   &NG@10 
   & 0.0450  &2.0438 &1.9730  &\textbf{2.3555}\\

    \midrule
   \multirow{2}{*}{HM}
   &HR@10
   & 0.0778 & \textbf{2.9385}  &1.8755 &2.7523\\
   &NG@10 
   & 0.0406 &\textbf{1.7457} &1.0095 &1.5740 \\
   \midrule
\multirow{2}{*}{Amazon}
   &HR@10
   & 0.1723 & 4.9190  &4.4885 &\textbf{5.6222}\\
   &NG@10 
   & 0.1381 &4.8547 &4.3779 &\textbf{5.4046} \\
   \bottomrule
    \end{tabular}}
    \end{center}
   \label{tb: cold}
\end{table}

\subsection{MMSR in Cold-start Scenario}  

In this section, we concentrate on the cold-start problem. MMSR is naturally suitable for such scenarios since their encoders are especially suitable for modeling the raw features of items, i.e., text and image, whether cold or not. 
To verify this, we conduct evaluations using IDSR, MMSR-T and MMSR-V, and MMSR in the cold-item scenario of the above \textbf{4} datasets.  

\textbf{Results.} Table \ref{tb: cold} reveals that MMSR, MMSR-T, and MMSR-V all notably outperform IDSR in predicting cold items. This superior performance stems from the fact that IDSR only derives representations from behavioral patterns, but MMSR incorporates original modal features, the advantage of which becomes more apparent for cold start. Interestingly, while MMSR-T and MMSR-V have comparable performances in standard situations (as is shown in Table \ref{tb: various_Rs}), MMSR-T tends to be more adept at handling cold items. This suggests that text-based representations might have a stronger generalizing capability for recommendation tasks than visual ones.

\subsection{Results of Transfer Learning}

We conduct the transfer learning experiments only on MMSR since IDSR is non-transferable.

\textbf{Scaling the source domain.} As shown in Table \ref{tb: hm_to_am}, we choose the HM-100K, -200K, and -500K datasets as the source domain and the e-commerce Amazon-10K \cite{amazon_1, amazon_2} dataset as the target domain. As the scale of the source domain increases, the results of transferring increase compared with those from scratch. This implies that by increasing the sample size, MMSR can more effectively uncover user preferences within the modal content.

\begin{table}[htbp]
   \caption{Transfer learning from HM to Amazon. \textbf{Source}: HM-100K, HM-200K, HM-500K. \textbf{Target}: Amazon-10K Clothes Dataset. ‘Improv.’ is the relative improvement of the best transferring compared to training from scratch.}
   \small
   \begin{center}
  \setlength{\tabcolsep}{3pt}{
   \begin{tabular}{lcccccc}
   \toprule
   \multirow{2}{*}{Metrics} & \multirow{2}{*}{IDSR}&\multirow{2}{*}{Scratch} & \multicolumn{3}{c}{Transfer} &\multirow{2}{*}{Improv.} \\
   \cmidrule(lr){4-6}
   & & &100K &200K & 500K & \\ 
   \midrule
   HR@10 & 20.2600 &25.4900  &26.3100 &26.2400 & \textbf{26.4900} & +3.80\% \\
   NG@10 & 17.2302 &21.7594 &23.3476 &22.9698 & \textbf{23.3558} & +6.83\% \\
   \bottomrule
\end{tabular}}
   \label{tb: hm_to_am}
\end{center}
\end{table}

\textbf{Scaling the target domain.} On the contrary, we select Kwai-200K as the source dataset and scale the target Bili dataset from 10K, 20K to 50K. As shown in Table \ref{tb:Kwai_to_bili}, the transferring improvement is reduced as the target domain expands.

\begin{table}[htbp]
   \caption{Transfer learning from Kwai to Bili. \textbf{Source}: Kwai-200K dataset. 
   \textbf{Target}: Bili-10K, -20K, and -50K dataset.  ‘Improv.’ is the relative improvement of the transferring compared to the corresponding training from scratch.}
   \begin{center}
   \small
  \setlength{\tabcolsep}{1.5pt}{
   \begin{tabular}{lcccccc}
   \toprule
   \multirow{2}{*}{} &\multicolumn{2}{c}{10K}  &\multicolumn{2}{c}{20K} &\multicolumn{2}{c}{50K}\\
   \cmidrule(lr){2-3}\cmidrule(lr){4-5} \cmidrule(lr){6-7}
     &HR@10 &NG@10 &HR@10 &NG@10 &HR@10 &NG@10\\ 
   \midrule	
   \multirow{1}{*}{ID} &1.0300 &0.5252 &1.3950 &0.7633 & 2.0420 &1.0727\\
   \midrule
   Scratch  & 1.4600 &0.7087 &2.0750  &1.1148 &2.7580 &1.3801\\
   Transfer & 1.5600 & 0.8101 &2.1600 &1.0587 &2.7760	&1.3689 \\
   \midrule
   Improv. & \textbf{+6.41\%} & \textbf{+14.31\%} &+3.93\% &-5.03\% & +0.64\% & -0.81\%\\
   \bottomrule
\end{tabular}}

   \label{tb:Kwai_to_bili}
\end{center}
\end{table}


\textbf{Results.} As is shown in Table~\ref{tb: hm_to_am} and Table~\ref{tb:Kwai_to_bili}, MMSR can perform better than training from scratch, but it does not always hold when the ratio of the source domain to the target domain is narrowed, e.g., the results of transferring from Kwai-200K to Bili-20K and -50K show no advantages. The performance drop might stem from differences in user groups between domains. Larger downstream datasets can introduce unrelated features and noise, hampering the transfer learning performances of sequential recommenders.

\begin{figure}[htbp]
	\begin{center}
    \includegraphics[width=1.0\linewidth]{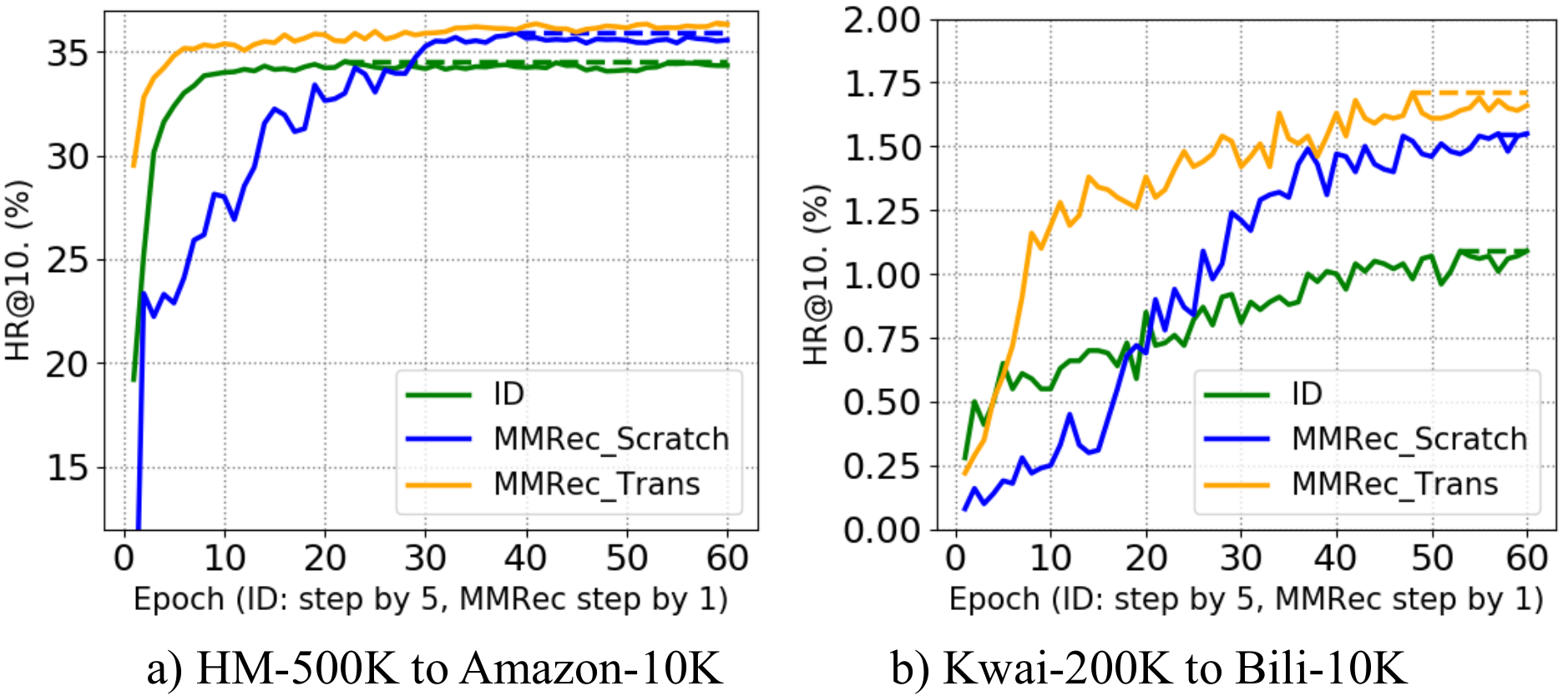}
    \end{center}
    \caption{Comparison of MMSR-Transfer and MMSR-Scratch.} 
    \label{Transfer_fig}
\end{figure}

\textbf{Convergence.} Figure \ref{Transfer_fig} shows transfer learning results for MMSR, using HM-500K and Kwai-200K as sources and Amazon-10K and Bili-10K as targets. IDSR was trained for 300 epochs, while MMSR-Transfer and MMSR-Scratch only needed 60 epochs. MMSR-Transfer demonstrated superior stability, faster convergence, and better performance compared to MMSR-Scratch.

\section{Bag of Tricks}
\label{Tricks}
In this section, we provide some practical insights and helpful tricks for training a performant MMSR. If not specified, we select a default setting of TE, VE, SR architecture, and dataset (i.e., RoBERTa, MAE, SASRec,  and Bili). 
\label{bag_of_tricks}

\subsection{Full Fine-tuning vs. Top-layer Fine-tuning}
\label{full-fine-tuning}
In all the above experiments, we fine-tune only the top blocks for all encoders by default, which is mainly due to computational consumption. Here, we try to investigate whether all parameters need to be adapted. Table \ref{tb:finetuning} shows the results of full and top-layer fine-tuning showed irregular fluctuations. Among them, top-layer fine-tuning works better in MMSR. We conclude that top-layer fine-tuning is worth considering in order to balance accuracy and efficiency in practice.

\begin{table}[htbp]
\caption{Full vs top-layer fine-tuning.} 
  \begin{center}
   \setlength{\tabcolsep}{3pt}{\begin{tabular}{lccccc}
  \toprule
  \multirow{2}{*}{-} & \multirow{2}{*}{Encoders} &\multicolumn{2}{c}{HM }  &\multicolumn{2}{c}{Bili}\\
  \cmidrule(lr){3-4}\cmidrule(lr){5-6}
   &  &HR@10 &NG@10 &HR@10 &NG@10 \\ 
  \midrule
  \multirow{1}{*}{ID}& - &8.4825 &5.2432 &3.1670 &1.7031 \\
  \midrule
  \multirow{2}{*}{MMSR-T}
  & RoBERTa   &8.2275  &4.6923  &\textbf{3.7790} &\textbf{1.9655}\\
   & RoBERTa-Full &\textbf{8.6835} &\textbf{5.0220} &3.7410 &1.9294 \\
  \midrule
  \multirow{4}{*}{MMSR-V}
  & MAE &6.1945 &3.2329 &2.4200 &1.1823 \\
  & MAE-Full &\textbf{7.3975} &\textbf{4.0094} &\textbf{3.1410} &\textbf{1.6026} \\
  \cmidrule(lr){2-6}
  & ResNet  &7.3520 &3.9049 &\textbf{3.2010} & \textbf{1.6545} \\
  & ResNet-Full &\textbf{7.3810} &\textbf{3.9253}  &3.1060  &1.6019 \\
  \midrule
  \multirow{2}{*}{MMSR}
  & Ro.+MAE	&\textbf{9.9635}	 &\textbf{5.7703} &\textbf{4.0810} &\textbf{2.1194} \\
  & Ro.-F+MAE-F &9.5235 	 &5.4953 &4.0500 &2.1063 \\
  \bottomrule
\end{tabular}}
  \label{tb:finetuning}
\end{center}
\end{table}

\subsection{Different Learning Rates}
As shown in Table \ref{tb:Learning_rate}, using the same learning rate for all modules of the model will lead to performance degradation, which may be because the pre-trained parameters already contain a certain amount of knowledge about vision and language, and over-tuning them may lead to loss of this valuable information.
As shown in Table \ref{tb:Learning_rate}, the performance of MMSR would deteriorate when all modules use the same learning rate. This is probably because the pre-trained parameters in the item encoders already contain a certain amount of knowledge about vision and language, and over-tuning may lead to the loss of this knowledge. But the downstream recommendation networks and fusion modules need training from scratch, so a large learning rate would be better.

\begin{table}[htbp]
  \caption{Comparison of MMSR with different learning rates for different modules. \textbf{Top}: TE and  VE of MMSR. \textbf{Downstream}: the recommendation networks and fusion modules.}
  \begin{center}
  \begin{tabular}{cccc}
  \toprule
  \multicolumn{2}{c}{Learning rate} &\multicolumn{2}{c}{MMSR} \\
  \cmidrule(lr){1-2}\cmidrule(lr){3-4}
  Top &Downstream &HR@10 &NDCG@10 \\ 
  \midrule
   5e-5 & 5e-5 & 3.8300 &1.9996\\
   8e-5 & 8e-5 & 4.0400 &2.1171\\
   1e-4 & 1e-4 &4.0810 &2.1194 \\
   8e-5 & 1e-4 & 4.0500 & 2.1096 \\
   5e-5 & 1e-4 &\textbf{4.0910} &\textbf{2.1331}\\
  \bottomrule
\end{tabular}
  \label{tb:Learning_rate}
\end{center}
\end{table}

\subsection{Length of Texts}
In our investigation, we dedicated substantial effort to analyzing the impact of varying quantities of textual information. This aspect is crucial, as the volume of textual data in a recommendation system can significantly influence its predictive accuracy. Our primary focus was on discerning how changes in the length of titles, an integral component of textual data, affected the performance of our models. As demonstrated in Table \ref{tb:text_tokens}, our exploration yielded noteworthy results. Both MMSR-T and MMSR exhibited a positive response to increases in title length. This trend implies that an expansion of title content can contribute to enhanced model performance. Our findings underline the significance of textual volume, particularly in title length, as a determinant factor for improving the functionality of MMSR.

\begin{table}[htbp]
  \caption{The impact of various lengths of text for MMSR-T and MMSR.}
  \begin{center}
  \begin{tabular}{lcccc}
  \toprule
  \multirow{2}{*}{Tokens} &\multicolumn{2}{c}{MMSR-T}  &\multicolumn{2}{c}{MMSR}\\
  \cmidrule(lr){2-3}\cmidrule(lr){4-5}
    &HR@10 &NDCG@10 &HR@10 &NDCG@10 \\ 
  \midrule
   10 &3.5360 &1.8597 &3.6710 &1.9008 \\
  30 &3.7650 &1.9455 &3.9810 &2.0765 \\
  50 &\textbf{3.7790}  &\textbf{1.9655} &\textbf{4.0810} &\textbf{2.1194}\\
  \bottomrule
\end{tabular}
  \label{tb:text_tokens}
\end{center}
\end{table}

\subsection{Mask Ratio of Images}
The volume of textual and visual information can greatly affect model performance MMSR. Inspired by MAE~\cite{he2022masked}, we control the image volume by adjusting its mask ratio.  Table \ref{tb:mask_ratios} shows that as the mask ratio increases, the performance of MMSR-V rises but MMSR decreases gradually. A possible reason is that information is insufficient for MMSR-V thus it would be better to keep more visual features. While for MMSR, the information may be saturated with both linguistic and visual features being fed, and it would be more intractable to optimize since the redundant image information would be taken as noise in this case.

\begin{table}[htbp]
\caption{The impact of the volume of image information for MMSR. Note that we use mask ratio from MAE~\cite{he2022masked} to control the volume of image information.}
   \begin{center}
   \begin{tabular}{lcccc}
   \toprule
   \multirow{2}{*}{Mask Ratio} &\multicolumn{2}{c}{Vision}  &\multicolumn{2}{c}{MMSR}\\
   \cmidrule(lr){2-3}\cmidrule(lr){4-5}
     &HR@10 &NDCG@10 &HR@10 &NDCG@10 \\ 
   \midrule
    0\% &\textbf{2.9130} &\textbf{1.4704} &3.7990 &1.9756 \\
    50\% &2.6500 &1.2560 &3.9780 &2.0996 \\
   75\% &2.4200 &1.1823 &\textbf{4.0810} &\textbf{2.1194}\\
   \bottomrule
   \end{tabular}
   \label{tb:mask_ratios}
\end{center}
\end{table}

\subsection{Visualization of IDSR and MMSR} 

In this study, we approach SR as a unique downstream application of multi-modal learning. Specifically, we employ solely multi-modal representations to ascertain whether they could match or even supersede traditional item-ID representations. We visualize both MMSR and IDSR from two perspectives of item and user. For this purpose, we selected the Bili dataset, randomly choosing 1000 items and 10 users, and then leveraged t-SNE \cite{van2008visualizing} tool to unveil their latent distributions.

\subsubsection{Visualization of Items}

\begin{figure}[htbp]
	\begin{center}
		\includegraphics[width=1\linewidth]{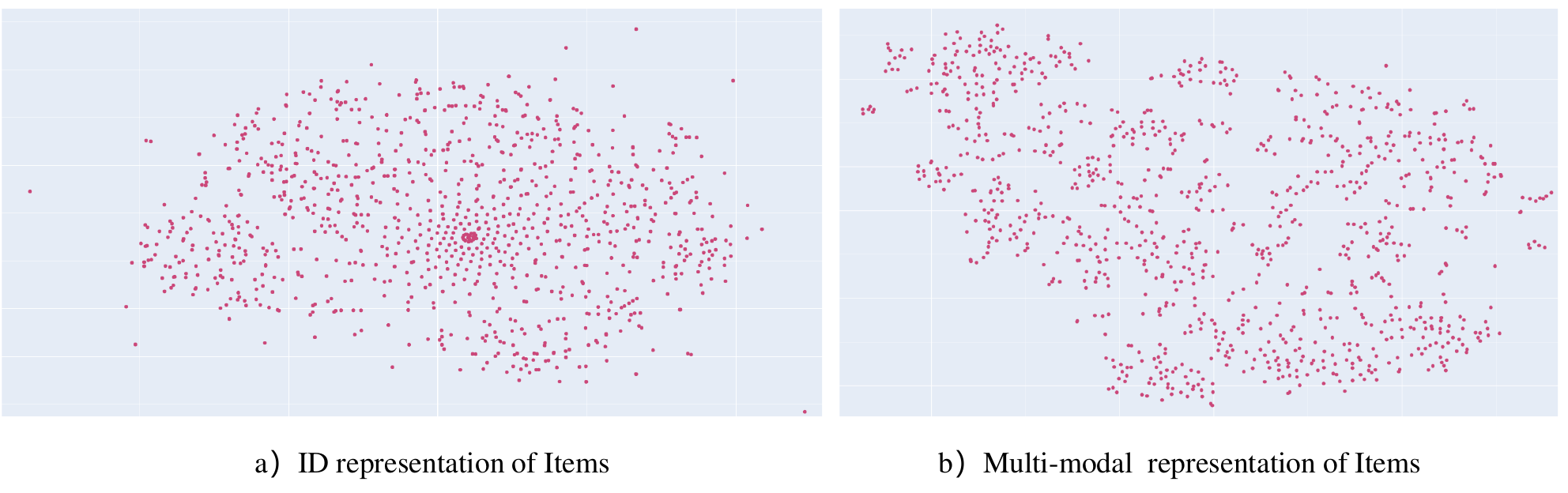}
	\end{center}
	\caption{Visualization of items in IDSR and MMSR.}
	\label{Item_visual}
\end{figure}

In our study, we analyzed item visualizations by item-ID, as presented in Figure \ref{Item_visual}, revealing two main issues: a clustering of popular items and the scattering of less interacted, 'cold' items, leading to a biased recommendation system favoring popular items. However, using multi-modal representations that include text and images leads to a more even distribution of items, reducing bias and improving the inclusion of 'cold' items, thus enhancing recommendation quality.



\subsubsection{Visualization of Users} 

\begin{figure}[htbp]
	\begin{center}
		\includegraphics[width=1\linewidth]{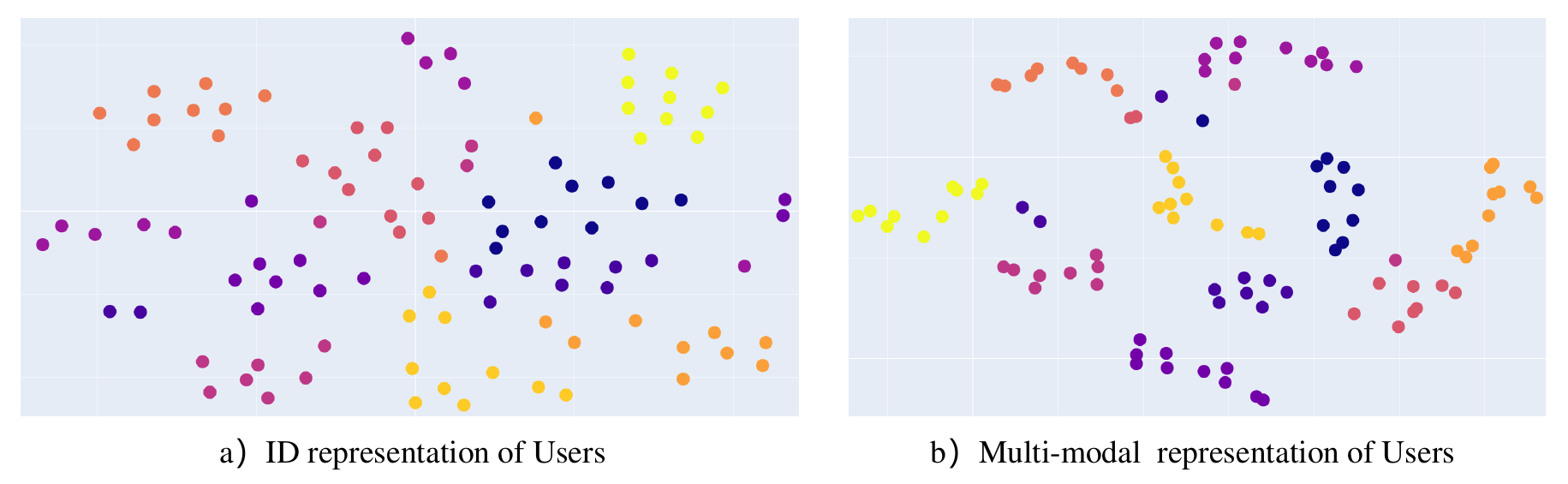}
	\end{center}
	\caption{Visualization of users in IDSR and MMSR}
	\label{User_visual}
\end{figure}

In our study, we compared user representations in IDSR and MMSR, focusing on their distribution within and across users. MMSR showed a closer clustering of items for each user, indicating its effectiveness in recognizing user behavior and preferences, especially for those with 'cold' items. Furthermore, MMSR preserved distinct distances between users, highlighting its ability to maintain individual user uniqueness and prevent preference homogenization. This is visually supported by Figure \ref{User_visual}, which shows clear clustering and inter-user distances.

\section{Conclusion}

In this paper, we propose a simple and universal Multi-Modal Sequential Recommender (MMSR) framework. We conduct comprehensive empirical studies along multiple dimensions, including text encoders, vision encoders, fusion strategies, multi-modal pre-training paradigms, and SR architectures. The experimental results demonstrate that MMSR not only completely outperforms the traditional IDSR, but also own decent transferability across platforms. Moreover, our work suggests an empirical potential for training specific multi-modal PTMs for unified SR.  As mentioned above, this paper is only a preliminary study of MMSR with the following limitations: (1) we only consider SR scenarios with text and visual modalities, not voice and video; (2) we only consider a very basic end-to-end way to introduce TE, VE, and VLPs into the recommendation model, without introducing any other technical tricks, the results we obtain from MMSR may still be suboptimal. To further serve RecSys communities, we plan to release a large-scale multi-modal RecSys dataset in the future, and specific multi-modal PTMs for unified SR frameworks.


\ifCLASSOPTIONcaptionsoff
\newpage
\fi
\bibliographystyle{IEEEtran}
\bibliography{refs}

\end{document}